\documentclass[FBSedit,FBSmath,ecsub]{FBSsuppl}

\usepackage{amsfonts}
\usepackage{amssymb}
\usepackage{graphicx}

\title{Comparison of Different Boost Transformations for the Calculation of
Form Factors in Relativistic Quantum Mechanics}
\author{Lukas Theu{\ss}l\instnr{1}, A. Amghar\instnr{2},
B. Desplanques\instnr{3}, S. Noguera\instnr{1}}
\instlist{Departamento de Fisica Teorica, Universidad de Valencia, 
E-46100 Burjassot (Valencia), Spain 
\and
Facult\'e des Sciences, Universit\'e de Boumerdes, 
35000 Boumerdes, Algeria 
\and 
Institut des Sciences Nucl\'eaires (UMR CNRS/IN2P3--UJF),  
F-38026 Gre\-noble Cedex, France}

\begin{document}

\maketitle

\begin{abstract}
The effect of different boost expressions, pertinent to the instant, 
front and point forms of relativistic quantum mechanics, 
is considered for the calculation 
of  the ground-state form factor of a two-body system in simple scalar models. 
Results with a Galilean boost as well as an explicitly covariant calculation 
based on the Bethe-Salpeter approach are given for comparison. 
It is  found that the present so-called point-form 
calculations of form factors strongly deviate from all the other ones.  
This suggests  that  the formalism which underlies them 
requires further elaboration. A proposition in this sense is made.
\end{abstract}


\section{Introduction}
\label{sec:intro}

Among the forms of relativistic quantum mechanics originally proposed by 
Dirac~\cite{Dirac:cp}, the point form is probably the least known and also the 
one least used. This may be due to the non-linear constraints on coordinates 
when
quantization is performed on a hyperbolo\"{\i}d. On the other hand, the dynamics
is particularly appealing in this form, because the interaction is solely contained
in the momentum operators, i.e., only four generators of the Poincar\'e group 
are dynamic, while, in particular, rotations and boosts are kinematic. 
Interest in the point form has been revived recently by the work 
of Klink~\cite{Klink:1998} and its applications to the form factors of the 
pion~\cite{Allen:hb}, the deuteron~\cite{Allen:2000ge} and the 
nucleon~\cite{Wagenbrunn:2000es}. 
In the deuteron case, there is no significant improvement with  
respect to a non-relativistic calculation. In the nucleon case, the  
agreement with experimental data is at first sight spectacular, 
especially at the lowest $Q^2$  range considered by the authors. 

However, analysing the calculations in detail, some questions arise. The
well-known vector-meson dominance mechanism, which explains to a large part the
nucleon form factors at low $Q^2$ via the coupling of the photon to the nucleon
by $\rho$ and $\omega$ exchanges, is not accounted for at all, and nothing
indicates that this might be achieved in a hidden way by incorporated
relativistic effects. Furthermore, no intrinsic quark form factors (which could
account for $\rho$ and $\omega$ exchanges implicitly) are employed, while they
are necessary for the construction of the quark-quark interaction.

From these works, one may deduce an expression for the mean squared radius
that scales like $1/M^2$, where $M$ is the total mass of the bound system. This
would have the surprising consequence that the root  mean squared radius
increases with decreasing $M$, and, in particular,
diverges in the limit $M\rightarrow 0$. This is contrary to physical intuition,
where a smaller total mass, usually brought about by increasing the interaction,
leads to a more compact system. Note that the limit $M\rightarrow 0$ is not a
completely academic one since it applies to the pion. A simple scaling argument
would lead to a squared radius of the pion of 
$3/m_\pi^2 \simeq 6 \, {\rm fm}^2$ (!).

The success of the calculation in ref.~\cite{Wagenbrunn:2000es} is attributed by
the authors to relativistic boost effects. Curiously, taking into account the
Lorentz-contraction effect by a simple replacement of the argument of the form 
factor, one gets an effect that increases the form factor for a given $Q^2$.
This is the opposite of what was found in ref.~\cite{Wagenbrunn:2000es}.

In a series of recent papers, motivated by the above mentioned observations, 
the reliability of the employed point-form implementation has been tested on 
some simple scalar systems, which are academic, but offer the advantage of 
eliminating uncertainties, like spin effects and intrinsic form factors of the 
constituents, while allowing partly analytic results. 
These works include a two-body system composed of scalar 
particles exchanging a zero-mass boson~\cite{Desplanques:2001zw} and a system 
corresponding to a zero-range interaction~\cite{Desplanques:2001ze}. A
systematic comparison with predictions of the other forms of dynamics (instant
and front form), as well as with a non-relativistic and an explicitly covariant
calculation based on the Bethe-Salpeter approach has been presented in 
ref.~\cite{Amghar:2002jx}, where also the
sensitivity of the results on the mass operator was investigated.

We shall only outline in this contribution the procedure that was followed in
ref.~\cite{Amghar:2002jx}, concentrating on the construction of the mass
operator, where another questionable feature of recent point form applications
arises. An example of numerical results for form factors will confirm the
unusual results obtained in the more realistic calculations of 
refs.~\cite{Allen:2000ge,Wagenbrunn:2000es}.


\section{Mass operator}
\label{sec:massop}

The starting point of every calculation is a dynamical equation (a mass
operator), from which a wave function may be determined that is subsequently
used for the calculation of observables. The important constraint to be
fulfilled by such an equation is its covariance, i.e., the resulting spectrum
should be invariant with respect to Lorentz boosts.
We choose an equation with a quadratic form of the energy- and the phase-space 
factors, which facilitates the calculations:
\begin{eqnarray} 
\nonumber  
\lefteqn{ 
\Big(E_P^2-(e_{p_1}+e_{p_2})^2\Big)\;\Phi(\vec{p}_1,\vec{p}_2)= 
\int\!\!\int \frac{d\vec{p}_1\!'}{(2\pi)^3}\;\frac{d\vec{p}_2\!'}{(2\pi)^3}}  
\\ && \qquad 
\frac{\sqrt{2\,(e_{p_1}+e_{p_2})}}{ \sqrt{2\,e_{p_1}} \, \sqrt{2\,e_{p_2}} }\, 
V_{int}(\vec{p}_1,\vec{p}_2,\vec{p}_1\!',\vec{p}_2\!')\; 
\frac{\sqrt{2\,(e_{p_1'}+e_{p_2'})}}{ \sqrt{2\,e_{p_1'}} \, \sqrt{2\,e_{p_2'}}  
}\; \Phi(\vec{p}_1\!',\vec{p}_2\!'), 
\label{20a} 
\end{eqnarray} 
where $E_P=\sqrt{M^2+\vec{P}^2}$, $e_p=\sqrt{m^2+\vec{p}\,^2}$. The 
quantities  $M$ and $\vec{P}$ represent the total mass and the total 
momentum of the system  under consideration.

One can easily determine the constraints that 
ensure the invariance of the mass spectrum of this equation. The same 
constraints also provide a direct way to relate a wave function 
calculated in a moving frame (with a finite  momentum $\vec{P}$), 
to the rest frame wave function. Provided  
$V_{int}(\vec{p}_1,\vec{p}_2,\vec{p}_1\!',\vec{p}_2\!')$ is appropriately 
chosen, the above equation can always be transformed into the center of mass by
a simple change of variables:
\begin{equation} 
(M^2-4\,e_{k}^2)\;\phi_0(\vec{k})= 
\int \frac{d\vec{k}'}{(2\pi)^3}\; 
\frac{1}{\sqrt{e_{k}}}\; V_{int}(\vec{k},\vec{k}')\; \frac{1}{\sqrt{e_{k'}}}\; 
\phi_0(\vec{k}'). 
\label{20c} 
\end{equation} 
This equation is of the form $M^2=M_0^2+V$, i.e., $M^2$
gives an invariant mass operator whose solutions can be used in
any form of relativistic quantum mechanics. It is just the relation between the
set of particle momenta, $\vec{p}_1$, $\vec{p}_2$, to the set of 'internal'
momenta,  $\vec{k}$, $\vec{P}$, that will be different from one form to the
other. In ref.~\cite{Amghar:2002jx} this transformation has been explicitly 
demonstrated for the instant form of dynamics. It is given by an expression that
is equivalent to the one obtained from the Bakamjian - Thomas 
construction~\cite{Bakamjian:1953kh}:
\begin{eqnarray}
\vec{p_1}&=&\vec{k}-\vec{P}\;\frac{\vec{k} \cdot \vec{P}}{P^2}
+\vec{P}\;\frac{\vec{k} \cdot \vec{P}}{P^2}\;\frac{\sqrt{4\,e_k^2+P^2}}{2\,e_k}
+e_k\;\frac{\vec{P}}{2\,e_k}, \nonumber \\
e_{p_1}&=&e_k\;\frac{\sqrt{4\,e_k^2+P^2}}{2\,e_k} 
+\vec{k} \cdot \frac{\vec{P}}{2\,e_k},
\label{20e}  
\end{eqnarray}
together with similar expressions for particle 2, where the 
change, $\vec{k} \rightarrow -\vec{k}$ has to be made.
In the instant form, the momenta of the constituent particles are 
related to the total  momentum $\vec{P}$ by the equality
$\vec{p}_1+\vec{p}_2=\vec{p}_1\!'+\vec{p}_2\!'=\vec{P}$, 
consistently with the property that the momentum has a kinematical 
character in this form. When the dynamics is described on a surface 
different from the instant-form one, $t=\tau$, other relations between 
momenta are obtained. This is easiest to see by integrating plane waves over the
hyper-surface under consideration. For a hyper-plane, $\lambda \cdot x=\tau$ 
with $\lambda^2 =1$, for instance, one obtains:
\begin{eqnarray}
\label{hplane}
\nonumber
\lefteqn{
\int d^4x \, \delta(\lambda\cdot x) \, e^{i(p_1+p_2-P)\cdot x} = } 
&& \\ && \qquad
(2\pi)^3 \frac{1}{\lambda^0}
\delta\left(\vec{p}_1+\vec{p}_2-\vec{P}-\frac{\vec{\lambda}}{\lambda^0}
(e_1+e_2-E_P)\right),
\end{eqnarray}
while for a hyperbolo\"{\i}d ($x^2=\tau$), one gets for the special case 
$\tau=0$:
\begin{equation}
\label{hpoloid}
\int d^4x \, \delta(x\cdot x) \, e^{i(p_1+p_2-P)\cdot x} = 
\frac{4\pi}{(\vec{p}_1+\vec{p}_2-\vec{P})^2-(e_1+e_2-E_P)^2+i\epsilon}.
\end{equation}
Eq.~(\ref{hplane}) would apply to an instant form of dynamics when
$\vec{\lambda}=0$, and we recover correctly $\vec{p}_1+\vec{p}_2=\vec{P}$ in
this case. However, in front form (Eq.~(\ref{hplane}) with 
$\vec{\lambda}/\lambda^0$ equal
to a unit vector $\vec{n}$) and in point form, Eq.~(\ref{hpoloid}), 
the relations
are more complicated. In particular, in neither of these forms we may set the
sum of the particle momenta ($\vec{p}_1+\vec{p}_2$) equal to zero in the rest
frame of the system ($\vec{P}=0$). This contrasts with the prescription used in
recent point form applications, where a relation of the form
\begin{equation}
\label{pfused}
 \vec{p}_1+\vec{p}_2 =  \frac{2e_k}{M} \,  \vec{P}
\end{equation}
is applied, i.e., $\vec{p}_1+\vec{p}_2 =0$ in the center of mass, which can
never be obtained from a relation of the form of eq.~(\ref{hpoloid}). The
relation between particle and 'internal' momenta obtained in that way rather 
resembles
the one in instant form given by Eq.~(\ref{20e}), with factors $2e_k$
replaced by the total mass $M$. This suggests, like  Eq.~(\ref{pfused}), that we
might expect troubles in the limit $M\rightarrow 0$ in this case.


\section{Results for elastic form factors}
\label{sec:results}

\begin{table}[htb!] 
\caption{Elastic vector- and scalar form factors, $F_1(Q^2)$ and  $F_0(Q^2)$,
for a system bound by the exchange of an infinite-mass boson (zero-range
interaction) for two values of the total mass $M$. 
The wave function used in the instant form (I.F.), front form (F.F.) and point
form (P.F.) cases is issued from Eq.~(\protect\ref{20c}). 
B.S. gives the results for a Bethe-Salpeter
calculation, while Gal. corresponds to a calculation employing a Galileian
boost. Asymptotic behaviors for $F_1(Q^2)$ are 
$Q^{-2}\;(\log Q)^2$,  $Q^{-2}\;(\log Q)^2$, $Q^{-4}$, 
$Q^{-2}\;(\log Q)^2$, and $Q^{-1}$  
for I.F., F.F., P.F., B.S. and Gal., respectively. 
\label{t30}} 
\medskip 
\begin{center} 
\begin{tabular}{cccccccc} 
\hline  \rule[0pt]{0pt}{3ex} 
  $Q^2/m^2$&             &   0.01   &   0.1  &  1.0   &  10.0 & 100.0 
  \\ [1.ex] \hline 
  \multicolumn{2}{c}{$M=1.6\,m$}                       &          &        &        &       &   \\ 
$F_1$&I.F.   & 0.999   & 0.990  & 0.917  &  0.594  & 0.208  \\ [0.ex]  
$F_0$&I.F.   & 1.325   & 1.309  & 1.176  &  0.658  & 0.187  \\ [1.ex]  
$F_1$&F.F.   & 0.999  & 0.989  & 0.908  &  0.566  & 0.191  \\ [0.ex]  
$F_0$&F.F.   & 1.325  & 1.309  & 1.176  &  0.659  & 0.187  \\ [1.ex]  
$F_1$&P.F. & 0.999   & 0.986  & 0.871  &  0.353 & 0.207-01 \\ [0.ex]	
$F_0$&P.F. & 0.998   & 0.976  & 0.800  &  0.236 & 0.108-01 \\ [1.ex]	
$F_1$&B.S.   & 0.999  & 0.989  & 0.908  &  0.566  & 0.191  \\ [0.ex]  
$F_0$&B.S.   & 1.325  & 1.309  & 1.176  &  0.659  & 0.187  \\ [1.ex]  
$F_1=F_0\;$&Gal.   & 0.999   & 0.994  & 0.947  &  0.699  & 0.320  \\ [1.ex] 
 \hline 
 \multicolumn{2}{c}{$M=0.1\,m$}                     &          &        &        &       &   \\ 
$F_1$&I.F.  & 0.999   & 0.996  & 0.963  &  0.759  & 0.343  \\ [0.ex]  
$F_0$&I.F.  & 1.498   & 1.487  & 1.389  &  0.920  & 0.320  \\ [1.ex]  
$F_1$&F.F.  & 0.999   & 0.995  & 0.954  &  0.723  & 0.315  \\ [0.ex]  
$F_0$&F.F.  & 1.498   & 1.487  & 1.389  &  0.920  & 0.320  \\ [1.ex]  
$F_1$&P.F. & 0.839  & 0.222   & 0.82-02 & 0.141-03 & 0.20-05 \\ [0.ex]    
$F_0$&P.F. & 0.699  & 0.130   & 0.42-02 & 0.071-03 & 0.10-05 \\ [1.ex]     
$F_1$&B.S.  & 0.999   & 0.995  & 0.954  &  0.723  & 0.315  \\ [0.ex]  
$F_0$&B.S.  & 1.498   & 1.487  & 1.389  &  0.920  & 0.320  \\ [1.ex]  
$F_1=F_0\;$&Gal.  & 0.999   & 0.998  & 0.980  &  0.846  & 0.475  \\ [1.ex] 
\hline \\ 
\end{tabular} 
\end{center} 
\end{table} 

Expressions for elastic charge form factors have been given in 
ref.~\cite{Amghar:2002jx} in the different forms of relativistic quantum
mechanics and for different choices of the two-body interaction model. We
reproduce in Table~\ref{t30} results obtained for a zero-range interaction 
(corresponding to an
infinite-mass exchange boson). This model has the advantage that some
calculations may be carried out analytically~\cite{Desplanques:2001ze}, 
which is useful for checking certain properties like the exact logarithmic 
dependence of the high $Q^2$ behavior of the form factors. The other extreme of
a zero mass exchange boson has been discussed in ref.~\cite{Desplanques:2001zw}.

Inspecting the results, two features are noticed immediately:
\begin{itemize}
\item[$\bullet$] The form factors in point form depart significantly from all 
the others, especially for high $Q^2$ and in the limit of small $M$.
\item[$\bullet$] The root mean squared radius of the bound state 
(proportional to the slope of the form factor at the origin) diverges in the 
limit $M\rightarrow 0$.
\end{itemize}
These features are qualitatively the same as the ones found in recent works,
where the so-called point form formalism was applied to the calculation of the
nucleon form factor~\cite{Wagenbrunn:2000es}.
In particular, because of the wrong power law behavior at high $Q^2$, 
the form factors in point form miss the Born amplitude, contrary to all the
other approaches. This is a severe shortcoming and one has to wonder about the
reasons for these peculiarities.

One possible explanation would be that two-body currents play a much
more dominant role in the point form than in all the other approaches. This
possibility is currently investigated~\cite{Theussl:2002}. In view of the 
unitary equivalence of all relativistic quantum mechanics approaches, and
because of the huge effect, one might wonder whether this is the most efficient
way to proceed.

An alternative to this approach was sketched in ref.~\cite{Desplanques:2001ze}.
It is based on the observation that, in practice, recent applications of the 
point-form~\cite{Allen:2000ge,Wagenbrunn:2000es,Desplanques:2001zw}
rely on employing wave functions  
issued from a mass operator whose solutions can also be identified 
with instant-form ones in the center of mass system. As noticed 
by Sokolov~\cite{Sokolov:1985jv}, this point-form approach is not 
identical to the one proposed by Dirac, where quantization is performed on 
a hyperbolo\"{\i}d, $x \cdot x=\tau$. 

When the system at rest described on the hyper-plane, $\lambda_0  
\cdot x=\tau$, is kinematically boosted to get initial and final states  
with four-momenta, $P_i^{\mu}$ and  $P_f^{\mu}$, these ones appear  
as described (quantized) on different surfaces,  $\lambda_i \cdot x=\tau$  
and  $\lambda_f \cdot x=\tau$, where $\lambda_{i,f}^{\mu} \propto  
P_{i,f}^{\mu}$ with $\lambda_{i,f}^2=1$. This feature results 
from the identification of point- and instant-form wave functions 
in the center of mass. It does not correspond to  the usual description 
of a process which, generally, relies on the same definition  
of the surface at all steps.

What is understood as point form in recent works is usually defined without any
reference to surfaces at all. This is not really in the spirit of Dirac, but it
is true that one does not need surfaces in order to construct a set of
generators that satisfies the Poincar\'e algebra. However,
the problem pointed out here concerns less the 
construction of the generators of the Poincar\'{e} algebra in 
terms of the total momentum $\vec{P}$ and the internal variable $\vec{k}$, but 
rather the relation of this set of variables to the physical ones. A point form
approach in the Dirac sense, with quantization performed on a hyperbolo\"{\i}d,
would at least resolve the issues of relations between variables,
that were raised in this paper.


\begin{acknowledge}
This work has been supported by the 
EC-IHP Network ESOP, under contract HPRN-CT-2000-00130 as well as the MCYT 
(Spain) under contract BFM2001-3563-C02-01.
\end{acknowledge}


\end{document}